\documentclass{ecai}
\usepackage{graphicx}
\usepackage{latexsym}

\usepackage{times}
\usepackage{soul}
\usepackage{url}
\usepackage[hidelinks]{hyperref}
\usepackage[utf8]{inputenc}
\usepackage{graphicx}
\usepackage{amsmath}
\usepackage{amsthm}
\usepackage{booktabs}
\usepackage{algorithm}
\usepackage{algorithmic}
\usepackage[switch]{lineno}
\usepackage{enumitem}   
\usepackage{multirow}
\usepackage{comment}
\usepackage{xcolor}
\usepackage{graphicx}
\usepackage{subfigure}

\newtheorem{assumption}{Assumption}

\newtheorem{definition}{Definition}

\ecaisubmission   % inserts page numbers. Use only for submission of paper.
\begin{document}

\begin{frontmatter}

\title{Combining Reinforcement Learning and Barrier Functions for Adaptive Risk Management in Portfolio Optimization}

% \author[]{\fnms{ }~\snm{Paper \#607}}
\author[A]{\fnms{Zhenglong}~\snm{Li}\thanks{Corresponding Author. Email: lzlong@hku.hk}}%\orcid{0000-0003-0305-9799}
\author[B]{\fnms{Hejun}~\snm{Huang}}%\orcid{0000-0003-3746-2105}}
\author[A]{\fnms{Vincent}~\snm{Tam}}%\orcid{0000-0003-4697-8277}} % use of \orcid{} is optional

\address[A]{Department of Electrical and Electronic Engineering, The University of Hong Kong, Hong Kong SAR, China}
\address[B]{Department of Aerospace Engineering, The University of Michigan, Ann Arbor, USA}

\begin{abstract}
% 200 words limit
Reinforcement learning (RL) based investment strategies have been widely adopted in portfolio management (PM) in recent years. Nevertheless, most RL-based approaches may often emphasize on pursuing returns while ignoring the risks of the underlying trading strategies that may potentially lead to great losses especially under high market volatility. Therefore, a risk-manageable PM investment framework integrating both RL and barrier functions (BF) is proposed to carefully balance the needs for high returns and acceptable risk exposure in PM applications. Up to our understanding, this work represents the first attempt to combine BF and RL for financial applications. While the involved RL approach may aggressively search for more profitable trading strategies, the BF-based risk controller will continuously monitor the market states to dynamically adjust the investment portfolio as a controllable measure for avoiding potential losses particularly in downtrend markets. Additionally, two adaptive mechanisms are provided to dynamically adjust the impact of risk controllers such that the proposed framework can be flexibly adapted to uptrend and downtrend markets. The empirical results of our proposed framework clearly reveal such advantages against most well-known RL-based approaches on real-world data sets. More importantly, our proposed framework shed lights on many possible directions for future investigation.

\end{abstract}

\end{frontmatter}

\section{Introduction}

In financial markets, only investing in a single asset brings huge uncertainties and risks once trading decisions are biased from the asset changes. To diversify investment risks, investors are suggested to allocate their capital to a set of assets with different natures during the trading period. However, as a fundamental financial problem, given a portfolio of financial products like stocks, futures, options and bonds, optimizing the ratios of assets in a portfolio to maximize returns at a low risk level is a challenge for all investors. According to the efficient market hypothesis \cite{fama1970efficient} and the investment market is an incomplete information game, there are numerous arbitrage opportunities existing in the financial market, but meanwhile they will be immediately filled in. Thus, accurately catching the change of assets by analyzing historical data is a key feature to construct profitable trading strategies in a portfolio. Inspired by the Modern Portfolio Theory \cite{markowitz2000mean}, more advanced portfolio theories such as Capital Growth Theory \cite{hakansson1995capital} and Black-Litterman Model \cite{black1992global} are presented to adapt the actual financial markets with practical constraints. Yet in the highly volatile financial markets, the traditional theories may not generate effective trading strategies due to many strict assumptions. Over the past decade, machine learning and deep learning techniques have been introduced to manage portfolios by predicting price movements \cite{wang2021hierarchical,duan2022factorvae} or directly optimizing the weights of assets \cite{yang2022smart,liang2021adaptive}. Through discovering underlying patterns from historical market data in both microeconomics and macroeconomics, those intelligent methods have achieved excess earnings than traditional methods in which the trading signals are generated by simple combinations with some handcrafted technical indicators. Furthermore, in terms of the mechanism that trading agents execute orders after interacting with financial markets, more efforts have been made recently on applying Reinforcement Learning (RL) to optimize online portfolios by observing the current states of the trading environment in real time \cite{xu2021relation,zhang2020cost,ye2020reinforcement}.  

However, most existing RL-based portfolio optimization methods may hardly learn an effective and stable trading strategy due to the data efficiency. More specifically, since the highly volatile financial market leads to the market style frequently changing, the trained RL agents may not achieve success on the real-time environment as the distribution of the current data may shift. This will surely increase the uncertainty of portfolios and also bring high-risk exposures. Besides, most previous RL-based methods aim to maximize the long-term profit yet less take into account the short-term risk management of a portfolio, whereas in fact that fund managers are more concerned about investment risk exposures than returns in volatile financial markets. Despite having potential returns, the risky investment may lead to a high maximum drawdown in a short period, which is unacceptable to capital holders. In addition, to balance the returns and risks, some combined performance indicators like sharpe ratio and sortino ratio integrating returns and risks are used as the optimization target, but they cannot explicitly manage portfolio risks in a single transaction. To constrain the system dynamics within safe regions on automatic driving and robotics, \cite{alshiekh2018safe,cheng2019end,huang2022barrier} introduce Barrier Function (BF) based constraint controllers to adjust decisions generated by model-free RL algorithms where any risky action will be compensated for maintaining safe states while RL agents keep exploring policies with high rewards. Yet \cite{cheng2019end} uses linear programming in simple cases, potentially suffering from complex constraints. \cite{huang2022barrier} employed sum-of-square programming to restrict the exploration of RL agents in polynomial systems, albeit at the expense of increased time costs. Furthermore, those RL-BF approaches do not satisfy the formulation of risk management in portfolio optimization. Moreover, the previous works strictly control the RL actions all the time for which they lack the flexibility to adapt to different scenarios.

To both explore profitable strategies and reduce risk exposures throughout the trading period, a Risk-manageable Portfolio Optimization (RiPO) framework integrating both RL-based trading agents and BF-based risk controllers is proposed in the paper to achieve high long-term profits under acceptable short-term risks. First, by formulating a portfolio management problem as a Partially Observable Markov Decision Process (POMDP), a model-free RL framework is given to explore profitable trading strategies. Second, a BF-based risk controller is constructed by the second-order cone programming in terms of risk constraints, monitoring the potential investment risks brought by aggressive RL trading strategies and then adjusting the portfolios for avoiding huge losses. In addition, considering the risk aversion of investors and different market states, two flexible mechanisms named Adaptive Risk Strategy (ARS) and Dynamic Contribution Mechanism (DCM) are proposed to adjust the strength of risk constraints and the impact of risk controllers to the overall trading strategies for adapting to different market styles. In uptrend markets, the proposed framework relaxes risk constraints to pursue higher excess returns under acceptable risk levels. Conversely, the risk exposure will be strictly constrained to avoid potentially huge losses in downtrend markets. This will enhance the flexibility of the proposed framework to invest assets in actual highly volatile financial markets. 

The main contributions of the proposed RiPO framework are summarized as follows: 
\begin{enumerate}
    \item The RiPO framework is the first attempt to integrate RL and BF-based constraint programming for financial applications. The risky trading decisions generated by RL agents can be continuously monitored and adjusted for explicitly managing the risk exposures while keeping the exploration ability of RL approaches to search for profitable strategies.
    \item Compared with the previous RL-BF methods only tested in simple cases, the risk controller of the proposed framework combines the second-order cone programming and BF-based constraints to formulate more complex applications in actual financial markets. By modeling the relationship between investment risks and acceptable risk ranges, the potential risks are effectively reduced particularly in downtrend markets.
    \item Instead of completely dominating RL agents by controllers the whole time, two adaptive mechanisms in the RiPO are described to flexibly adjust the impact of risk controllers in terms of investor preference and market states, which earns higher returns by loosing risk constraints in uptrend markets but strictly manages risks in downtrend markets for reducing losses.
\end{enumerate}

Due to the nature of financial markets, it should be pointed out that portfolio risk management is not an absolute control that manages risks under any expected level in any case. In fact, the proposed framework is expected to avoid risky investments as possible so that the maximum drawdown and overall losses can be reduced.

\section{Preliminaries}

\subsection{Portfolio Optimization}

Online portfolio management is a multi-period trading strategy that the capital is periodically reallocated to the selected assets. In this work, there are two assumptions listed below.

\begin{assumption}\label{asm:1}
The portfolio will be only considered from long positions in this work.
\end{assumption}

\begin{assumption}\label{asm:2}
The turnover rate of assets in a portfolio satisfies the requirements of each order execution.
\end{assumption}

\noindent
Assumption \ref{asm:1} implies that investors cannot short assets unless they hold the long positions, while Assumption \ref{asm:2} encourages the evaluation of the proposed framework more close to reality. Based on these considerations, two primary objectives in portfolio optimization task are given, return maximization and risk minimization, respectively. Some basic financial terms are introduced as follows: 

\begin{definition}\label{def:portfolio}
(Portfolio Value) The value of a portfolio at time $t$ can be denoted by 
\begin{equation}
    C_t = \sum_{i=1}^{N} w_{t,i} p^{c}_{t,i},
\end{equation}
where $N$ is the number of assets in a portfolio, $w_{t,i}$ is the weight of an $i^{\text{th}}$ asset, and $p^{c}_{t,i}$ is the close price of $i^{\text{th}}$ asset at time $t$. Therefore, the portfolio is constrained based on Assumption \ref{asm:1} and \ref{asm:2} as 
 \begin{equation}
\forall w_{t,i}\in \mathbf{W}_t:\quad w_{t,i}\geq 0 ,\sum_{i=1}^N w_{t,i}=1,
\end{equation}
where $\mathbf{W}_t\in \mathbf{W}$ is the weight vector $\mathbf{W}$ at time $t$. 
\end{definition}

Definition \ref{def:portfolio} implies that the risk would be varied in terms of purposes, the corresponding covariance-weight risk from the Markowitz model \cite{markowitz2000mean} and the volatility of strategies provided a view of the short-term risk and long-term risk.

\begin{definition}
\label{def:shorttermrisk}
 (Short-term Risk) The portfolio risk at time $t$ can be presented as below 
 \begin{equation}
    \begin{aligned}
    \sigma_{p,t} & = \sigma_\beta + \sigma_{\alpha,t}\\
    \sigma_{\alpha,t} & = \sqrt{\mathbf{W}^T_{t}\Sigma_k \mathbf{W}_{t}} = \Vert \Sigma_k \mathbf{W}_{t} \Vert_2.
    \end{aligned}
\end{equation}
\noindent
where $\sigma_{\alpha,t}$ is the trading strategy risk, $\sigma_\beta$ is the market risk and $\mathbf{W}_{t}\in \mathcal{R}^{N \times 1}$ is the matrix of weights. The covariance matrix $\Sigma_k\in\mathcal{R}^{N\times N}$ between any two assets can be calculated by the rate of daily returns of assets in the past $k$ days. 
\end{definition}

\begin{definition}\label{def:longtermrisk}
(Long-term Risk) The strategy volatility is used to measure the portfolio risk in the whole trading period, which is the sample variance of daily return rate of the trading strategy. 
\end{definition}

\begin{definition}\label{def:sharperatio}
(Sharpe Ratio) The Sharpe Ratio (SR) is a usual performance indicator for evaluating a portfolio with the consideration of returns $R$, risk-free rate $r_f$ and portfolio risk $\sigma$, which is given as
 \begin{equation}
\text{SR} = \frac{R-r_f}{\sigma}.
\end{equation}
\end{definition}

Portfolio optimization has been studied in few decades. The technical analysis methods can be concluded into four categories including Follow-the-Winner, Follow-the-Loser, Pattern Matching Approaches, and Meta-Learning Algorithms \cite{li2014online}. They try to capture the price momentum by using handcrafted financial indicators. Recently, more investors are attracted by DL/RL technique. Except for the regular price data, \cite{liang2021adaptive,ye2020reinforcement} introduce news data to collect extra information for the portfolio management. In terms of model structures, \cite{xu2021relation,jiang2017deep} present specific modules to deal with assets information independently and also capture the correlations among assets. In addition, \cite{yang2022smart} adjusts portfolios and optimizes trading time points to achieve the online trading in minute levels. Nevertheless, most of the studies on portfolio optimization cannot explicitly constrain the investment risk exposure when using RL-based approaches to explore profitable strategies.

\subsection{Barrier Function}
Originally inspired from Lyapunov functions, barrier function is introduced to identify safe regions and drive controllers working inside the defined safe boundaries in control theory \cite{ames2019control,ames2014control}. Assume that a system dynamic can be denoted as
\begin{equation}
    s_{t+1} = f(s_t)+g(s_t)a_t+d(s_t),
\end{equation}
\noindent
where $s_t\in S$ is the system state at $t$, $a\in A$ is the action at $t$, $f:S\rightarrow S$ is the nominal unactuated dynamics, $g:S\rightarrow A$ is the nominal actuated dynamics and $d:S\rightarrow S$ is the unknown dynamics. A safe set 
\begin{equation}
C=\{s\in S: h(s,a)\geq 0\},
\end{equation}
can be described by the superlevel set of a barrier function $h:S\rightarrow \mathbf{R}$ in this dynamical system, where $h$ is a continuously differentiable function and also satisfies that $\frac{\partial f}{\partial s}\neq 0$ when $h(s)=0$. According to Nagumo's Theorem \cite{blanchini1999set}, the safe set $C$ will be forward invariant if there exists
\begin{equation}
\forall s\in C,\quad \frac{\Delta h(s_t,a_t)}{\Delta t} \geq 0 ,
\end{equation}
\noindent
where $\Delta t$ represents a time interval, and $\Delta h(s_t,a_t) = h(s_{t+1})-h(s_t)$ when considering a discrete-time barrier function. Further, considering the relaxation for safe constraints with a locally Lipschitz class $\mathcal{K}$ function $K$ such that 
\begin{equation}
     \mathop{\sup}\limits_{a_t\in A}\; [h(s_{t+1})-h(s_t)+K(h(s_t))] \geq 0,
\end{equation}
If there exists a feasible action $a_t$ satisfying the above BF-based constraint, then the system can be expected stay at the safe state at time $t+1$. This can fill the gap that the explored actions generated by RL agents may not concern the status of each state due to the long-term reward expectation. 

\section{Problem Formulation}
Since financial markets are influenced by many factors like unpredictable black swan events and system risks, it is difficult to collect all relevant information for a perfect investment decision. Therefore, instead of directly capturing the actual market states that are the hidden states of financial markets, the trading strategies can only rely on part of observable market data. Typically, the meta observable market data are the historical prices and volumes of each asset in a portfolio.

\subsection{Partially Observable Markov Decision Process}
For simplifying the optimization process, it is assumed that the next actual market state $s_{t+1}$ solely depends on the current actual market state $s_t$ as: $p(s_{t+1}| s_t,s_{t-1},\dots,s_1)=p(s_{t+1}|s_t),s\in S$, where $p$ is the conditional probability and $S$ is a finite set of actual states. Besides, the set of meta observable state of $i^{\text{th}}$ asset at timestamp $t$ can be denoted as $o_{t,i}^{\text{meta}}=\left\{  p^o_{t,i},p^h_{t,i},p^l_{t,i},p^c_{t,i},\text{vol}_{t,i} \right\}$, where $p^o_{t,i},p^h_{t,i},p^l_{t,i},p^c_{t,i}$ are the open/high/low/close price, and $\text{vol}_{t,i}$ is the trading volume. Furthermore, except for the meta observable states, some extra technical indicators derived from the $o_{t,i}^{\text{meta}}$ will be introduced to be a part of observable market states to help analyze underlying patterns of market trends. 

Define $o^{tech}_{t,i}=\left\{ k_{t,i,1},k_{t,i,2},\dots,k_{t,i,j} \right\}$, where $o^{tech}_{t,i}$ is a set of technical indicators, and $k_{t,i,j}$ is the $j^{\text{th}}$ technical indicator of $i^{\text{th}}$ asset at timestamp $t$. Beyond that, the current account status can be observed and be considered to make reasonable trading signals, which can be denoted as $o_t^a = \log \frac{C_t}{C_{\text{init}}}$, where $C_t$ is the current capital and $C_{\text{init}}$ is the initial capital.

In general, the portfolio management process can be modeled as a POMDP that can be defined as a tuple $\left(S,A,T,R,\Omega,O,\gamma\right)$, where $S$ denotes a finite set of actual market states, $A$ is a finite set of actions, $T(s_{t+1}|s_t, a_t) $ denotes a set of conditional transition probabilities between $s_{t+1}$ and $s_t$ under the action $a_t$, $R(s_{t+1}|s_t,a_t) $ presents the reward function, $\Omega$ indicates a finite set of observable states, $O(o_{t+1}|s_{t+1},a_t)$ is a finite set of conditional observation probabilities between $o_{t+1}$ and $s_{t+1}$ under the action $a_t$, and $\gamma\in[0,1)$ is the discount factor. The objective of portfolio management is to learn the decision policy $\pi:S\rightarrow A$ that can maximize the expected total rewards at all timestamps. It can be defined as  
\begin{equation}
    J(\pi^*) = \mathop{\max}\limits_{\pi\in \Pi}\; \mathbf{E}_{\tau\sim \pi}[\sum_{t=1}^{\infty}\gamma^{t-1}R_t],
\end{equation}
where $\pi^*$ is the optimal policy, $\tau\sim\pi$ is a trajectory under the policy $\pi$ and $\Pi$ is the possible policy space. To further approximate the solutions of POMDP problems, the history of previous observations up to the current timestamp can be recognized as a pseudo-state to estimate the actual state through a mapping function $\phi:\mathcal{H}\rightarrow\phi(\mathcal{H})$, where $ \phi\left(\mathcal{H}\right)={\phi\left(H\right)|H\in\mathcal{H}}$, $H_t\in\mathcal{H}$ is the observation history up to timestamp $t$, $\mathcal{H}$ is the space of all possible observable histories \cite{franccois2019overfitting}. 

Furthermore, the POMDP problem can be reformulated as a tuple $(\hat{S},A,\hat{T},\hat{R},\gamma)$. Specifically, $\hat{S}=\phi(\mathcal{H})$ is the estimated actual states, $\hat{T}({\hat{s}}_{t+1}|{\hat{s}}_t,a_t)$ is the estimated transition function in which ${\hat{s}}_{t+1},{\hat{s}}_t\in\hat{S}$ and $a_t\in A$, $\hat{R}({\hat{s}}_{t+1}|{\hat{s}}_t,a_t)$ is the estimated reward function, and the decision policy $\pi:\ \phi\left(\mathcal{H}\right)\rightarrow A$ with $\pi\in\Pi$. Accordingly, the bellman equation should be reformulated as
\begin{equation}
\begin{aligned}
V^\pi\left({\hat{s}}_t\right)=&R\left({\hat{s}}_t,\ \pi\left({\hat{s}}_t\right)\right)+\\
&\gamma\sum_{{\hat{s}}_{t+1}}{P\left({\hat{s}}_{t+1}\middle|{\hat{s}}_t,\pi\left({\hat{s}}_t\right)\right)V^\pi({\hat{s}}_{t+1})},
\end{aligned}
\end{equation}
where ${\hat{s}}_t=\phi(H_t)$, $V^\pi\left({\hat{s}}_t\right)$ is the expected reward in ${\hat{s}}_t$ under the policy $\pi$.

\subsection{Observation and Action}
Another featured property of portfolio optimization is that the trading signals given by the agent will not significantly influence the trend of asset prices unless the trading volumes are large enough and the paper will not discuss such extreme cases. Thus, the market observation function can be reformulated as 
\begin{equation}
O(o_{t+1}^{meta},o_{t+1}^{tech}|s_{t+1},a_t)\approx O(o_{t+1}^{meta},o_{t+1}^{tech}|s_{t+1}),
\end{equation}

As discussed in \cite{liu2020adaptive}, the conditional transition probability between account status $o_{t+1}^a$ and $a_t$ can be expressed as $O(o_{t+1}^a|a_t)$. Then the observation transition probability can be reformulated as $O(o_{t+1}|s_{t+1},a_t )$, where $o_{t+1}=({o_{t+1}^{meta},o_{t+1}^{tech},o_{t+1}^a})$ and $o_{t+1}\in\Omega$. Related technical indicators are listed in Appendix.

Since the financial markets in different countries have different regulations. To simplify the trading behaviors, only the long position is considered in this paper. The normalized weight of $i^{\text{th}}$ asset in a portfolio is defined as $a_{t,i}$, where $a_{t,i}\in[0,\ 1]$, $\sum_{i=1}^{N}a_{t,i}=1$, and $N$ is the number of assets in the portfolio. 

To close the realistic trading environment, two practical factors including transaction cost $\varsigma$ and slippage $\xi$ are considered in each transaction. The reward at timestamp $t$ can be defined as
\begin{equation}
    r_t=[-\varsigma +\sum_{i=1}^N a_{t,i}(\frac{p_t^c-p_{t-1}^c+\xi}{p_{t-1}^c})]\eta,
\end{equation}
where $\xi \sim \mathcal{U}\left(-\xi_{lower},\ \xi_{upper}\right)$, $\xi_{\text{lower}}$ and $\xi_{\text{upper}}$ are the lower and upper boundaries of slippage, and $\eta$ is the scaling factor in the reward function.

\section{Methodology }

\subsection{Overall Framework }

\begin{figure*}[ht] 
    \centering
    \includegraphics[width=14cm]{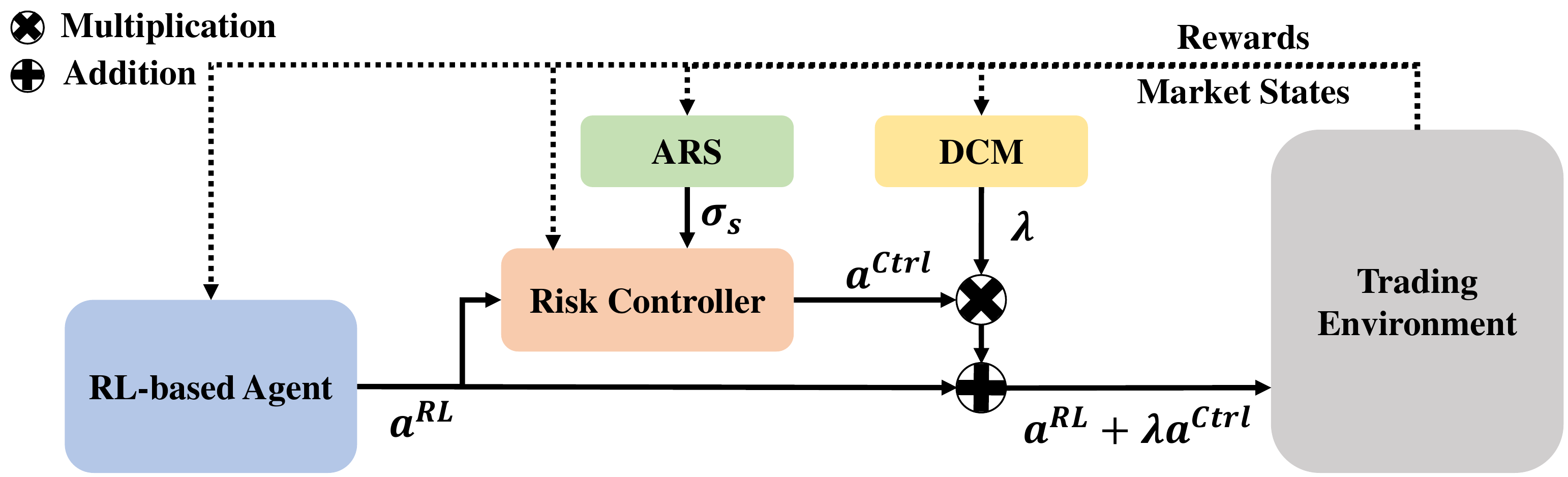}
    \caption{The Overview of RiPO.}% The results of ERM are from~\cite{iwasawa2021T3A}.}
    \label{fig: workflow}
\end{figure*}

An overview of the RiPO is depicted in Fig. \ref{fig: workflow}. The final investment strategy in the RiPO framework comes from the RL-based trading agent and risk management module. More specifically, the risk management module includes a BF-based risk controller, a Dynamic Contribution Mechanism (DCM), and an Adaptive Risk Strategy (ARS). Initially, learned from the current market states and the optimized policy, the RL-based trading agent suggests the weights of assets in a portfolio for the next trading period. However, some suggestions may ignore short-term risks as the RL-based trading agents are expected to earn long-term profits. To balance the expected long-term returns and short-term risks, the risk controller evaluates the risk exposure of original RL-based trading strategies and dynamically adjusts the portfolio for managing the near future risk within an acceptable range. Since the financial market always changes, the risk controller should adapt to different market states for higher returns and lower risks. Thus, there are two adaptive mechanisms named DCM and ARS enhancing the flexibility of risk controller to monitor the RL-based trading agents from the perspective of the impact of risk controllers and the strength of risk constraints. The detailed steps of the RiPO framework are described in Algorithm \ref{alg:algorithm}.

\begin{algorithm}[tb]
    \caption{The Training Procedure of the RiPO Framework}
    \label{alg:algorithm}
    \textbf{Input}: RL algorithm settings, portfolio trading settings \\
    \textbf{Output}: The optimal RL policy $\pi^*$
    \begin{algorithmic}[1] %[1] enables line numbers
        \STATE Initialize the RL policy $\pi_0$ and memory tuple $\hat{D}$.
        \FOR{$k=1$ to $Episode$}
        \STATE Reset the trading environment and set the initial action $a_0$.
        \FOR{$t=1$ to $T$}
        \STATE Observe the current market state $o_t$.
        \STATE Calculate the reward $r$.
        \STATE Store tuple ($o_{t-1}$, $a_{t-1}$, $o_t$, $r_{t-1}$) in $\hat{D}$.
        \STATE Sample suggested action $a_t^{RL}$ from the RL-based agent in terms of the current policy $\pi$.
        \STATE Update acceptable risk $\sigma_{s,t}$ by using the ARS module.
        \STATE Collect adjusted action $a_t^{Ctrl}$ from the risk controller.
        \STATE Update the contribution factor of the risk controller $\lambda_t$ .
        \STATE Adjust the current portfolio with the action $a_t=a_t^{RL}+\lambda_t a_t^{Ctrl}$.
        
        \IF {the RL policy update condition is triggered}
        \STATE Update the RL policy $\pi$ by learning the historical trading data from $\hat{D}$.
        \ENDIF

        \ENDFOR
        \ENDFOR
        \STATE \textbf{return} the optimal RL policy $\pi^*$
    \end{algorithmic}
\end{algorithm}

\subsection{Barrier Function-based Risk Management with Reinforcement Learning}

The RL-based portfolio optimization approach has great exploration capabilities on discovering profitable strategies, but it may give unreasonable actions when the current data distribution shifts due to financial market changing. Conversely, the programming-based methods can strictly satisfy the required constraints yet lack the abilities to explore the underlying patterns from raw data. Given by that, with the integration of barrier function to constrain the portfolio risk exposure, a model-based risk controller is formulated to cooperate with RL-based trading agents for modeling online portfolio optimization problem concerning both long-term returns and short-term risks. First, the system dynamics can be written as 
\begin{equation}
    \left[\begin{matrix}{\dot{P}}_{t+1}^c\\\sigma_{p,t+1}\\\end{matrix}\right]=\begin{bmatrix} \Delta P_{t+1}\\ \sigma_{\beta}\end{bmatrix} +\begin{bmatrix} 0\\ \Sigma_{k,t+1}^{\frac{1}{2}}\end{bmatrix}a_t,
\end{equation}
where $a_t=a_t^{RL}+a_t^{Ctrl}$ represents the upcoming adjusted weight of assets at time $t$, $\Sigma_{k,t+1}$ is the covariance matrix, $\sigma_{p,t}$, $\sigma_\beta$ are the portfolio risk and market risk as denoted in Definition \ref{def:shorttermrisk}. In terms of the risk-aware investment intuition, the objective of risk controllers is reducing the loss of expected profits while satisfying the risk constraint. Thus, the risk controller can be modeled as 
\begin{equation}
\begin{aligned}\label{eq:riskctrl}
	& && a_t^{Ctrl}=\mathop{\arg\min}\limits_{a_t^{Ctrl}}\;{\sum_{i=1}^{N}{-a_{t,i}^{Ctrl}\Delta p_{t+1,i}}}\\
	&\text{s.t.} &&h\left(\sigma_{p,t+1}\right)-h\left(\sigma_{p,t}\right)+\alpha\left(h\left(\sigma_{p,t}\right)\right)\geq0,\\
	& && 0\le a_{t,i}^{RL}+a_{t,i}^{Ctrl}\le1,\ \forall i\in{1,\ 2,\ ..,\ N},\\
        & && \sum_{i=1}^{N}{{(a}_{t,i}^{RL}+a_{t,i}^{Ctrl})}=1,
\end{aligned}
\end{equation}
\noindent
where $a_t^{Ctrl}$, $a_t^{RL}\in\mathbf{R}^N$, $\Delta p_{t+1}$ is the estimated price change from t to t+1 in terms of the moving average idea in this paper. For the risk constraint $\sigma_{p,t}\in[0,\sigma_{s,t}]$, there exists an acceptable set C such that
\begin{equation}
    C = \{\sigma_{p,t}:h\left(\sigma_{p,t}\right)\geq0 \},
\end{equation}
\noindent
where $\sigma_{s,t}$ is the upper boundary of acceptable risk at $t$. Then, the portfolio risk can be managed within an acceptable region if it satisfies 
\begin{equation}
     \mathop{\sup}\limits_{a_t\in A}\; [h\left(\sigma_{p,t+1},\ a_t\right)-h\left(\sigma_{p,t}\right)+K(h(\sigma_{p,t}))] \geq 0.
\end{equation}

Define $h=p^T s+q$, ($p \in\mathbf{R}^n, q\in\mathbf{R}$), let 
$$
c^*=h\prime(\sigma_{s,t+1},\sigma_\beta)-h\left(\sigma_{p,t}\right)+\alpha(h(\sigma_{p,t})).
$$ 

Furthermore, considering the portfolio risk constraint $0 \leq \sigma_{p, t} \leq \sigma_{s, t}$, the barrier function $h$ can be redefined as

     \begin{equation}
        \begin{aligned}
        h\left(\sigma_{p, t}\right)=\sigma_{s, t}-\sigma_{p, t}.
        \end{aligned}
    \end{equation}

The BF-based constraint can be reformulated as below 
\begin{equation}
\begin{aligned}
        \sigma_{\alpha,t+1}&=\sqrt{a_t^T\ \Sigma_{k,t+1}a_t}\\
        &=\Vert \Sigma_{k,t+1}^{\frac{1}{2}} a_t\Vert_ 2 \\
        &= \Vert \Sigma_{k,t+1}^{\frac{1}{2}} a_t^{Ctrl}+\Sigma_{k,t+1}^{\frac{1}{2}}  a_t^{RL}\Vert_ 2 \\
        & \leq c^*,
\end{aligned}
\end{equation}
\noindent
where $\Sigma_{k,t+1}$ is calculated by the historical price series and the estimation of $\Delta p_{t+1}$, and $\Sigma_{k,t+1}^{\frac{1}{2}}a_t^{RL}$ is a constant. Thus, the risk controller performs as a second-order cone program. More detailed derivation is given in the Appendix. 

After collecting the compensating adjustment $a_t^{Ctrl}$ of portfolios from the BF-based controller, the final trading decision $a_t$ would be made for the next trading period. The rewards $r_t$ and its action $a_t$ are stored into the memory of RL algorithms for further training, which can promote RL training efficiency for reaching optimal policy. 
% \begin{equation}
%  J(\pi^*) = \mathop{\max}\limits_{\pi\in \Pi}\; \mathbf{E}_{\tau\sim \pi}[\sum_{t=1}^{\infty}\gamma^{t-1}R_t].
% \end{equation}
% The RL policy optimization will contain the actions from the RL agent itself and also the actions from BF-based risk controller during the learning process.

\subsection{Dynamic Contribution Mechanism}

Intuitively, the risk preferences of investors are not fixed the whole trading period in actual financial markets. Particularly, they are willing to take higher investment risks for gaining higher returns in uptrend markets. Therefore, more aggressive and risky investment strategies are allowed at the moment. Conversely, when a downtrend market appears, most investors prefer to strictly constrain the portfolio risk to avoid huge losses even though there will miss some profitable opportunities. Originally inspired by \cite{stanovov2021nl} using non-linear transformation to balance exploration and exploitation at different optimization stages, a dynamic contribution mechanism is introduced to adaptively regulate the impact of risk controller at each transaction by considering the strategy risk exposures and investor preferences, which balances the exploration in RL-based agents and exploitation in risk management. To be more specific, according to the near performance of trading strategies, a scaling factor $\lambda_t\in[0,1]$ is given by a non-linear transformation to update the contributions of risk controllers to the final trading signals. The trading strategy with greater losses will be subject to tighter risk constraints to avoid aggressive investment decisions from RL-based agents.

\begin{equation}
    \lambda_t =\left\{\begin{aligned}
    &( m+G)^{(1-G)}, &&\quad R_s-r_f <0,\\
    &m, &&\quad \text{Otherwise},\\
    \end{aligned}\right.
\end{equation}
\noindent
where $G=\min(\frac{|R_s-r_f|}{v}, 1)$, $m\in[0,1]$ denotes the minimal impact of risk controllers to the overall trading strategy. The higher $m$ represents more strict risk management in which the $\lambda_{t}$ will be larger at the same loss of strategies. Note that, the risk requirements are strictly constrained at the whole trading period when $m=1$ such that $\lambda=1$. $v\in(0,1]$ represents the risk appetite of investors. In terms of qualitative analysis, the lower $v (v\rightarrow 0)$ has less tolerance on investment risk. It means that the small losses in a short-term period will trigger strict risk control. $R_s$ is the recen performance of trading strategies. Specifically, $R_s$ is given by the moving average of daily returns of trading strategies in this paper. Furthermore, the final trading decision at time t is revised as 
\begin{equation}
a_t=a_t^{RL}+\lambda_t a_t^{Ctrl}.
\end{equation}
\noindent 
where $a^{RL}_{t}$ satisfies $\sum_{i=1}^{N} a^{RL}_{t,i}=1$ and $a^{Ctrl}_{t}$ satisfies $\sum_{i=1}^{N} a^{Ctrl}_{t,i}=0$. Thus, the sum of $a_t$ adds up to 1. 

\subsection{Adaptive Risk Strategy }
As the strict risk management would lead to trading agents missing some potentially profitable opportunities in uptrend markets, the strength of risk constraint in the BF-based controller should be dynamic in terms of the risk preferences of investors and current financial market states in which investors expect lower investment risks in downtrend markets while allowing relatively high risks in uptrend markets for earning higher profits. Therefore, a simply yet efficiently adaptive risk strategy is introduced to enhance the adaptability of the proposed framework to the actual financial market. Considering the balance between expected returns and acceptable risks, the adaptive risk upper boundary $\sigma_{s,t+1}$ for the BF-based risk constraint is shown below.
\begin{equation}
    \sigma_{s, t+1} = \left\{\begin{aligned}
    &\sigma_{s, \min }, &&\quad \bar{R}_{t+1} \in\left(-\infty,(1-\mu) r_{f}\right), \\
    & M \bar{R}_{t+1}+b, &&\bar{R}_{t+1} \in\left[(1-\mu) r_{f},(1+\mu) r_{f}\right], \\
    & \sigma_{s, \max }, &&\quad \bar{R}_{t+1} \in\left((1+\mu) r_{f},+\infty\right),
    \end{aligned}\right.
\end{equation}
\noindent
where $M=\frac{\sigma_{s, \max }-\sigma_{s, \min }}{2 \mu r_{f}}$, $b=\frac{(1+\mu) \sigma_{s, \min }-(1-\mu) \sigma_{s, \max }}{2 \mu}$, $\bar{R}_{t+1}$ is the expected returns, $\sigma_{s,min}$ and $\sigma_{s,max}$ are the minimum and maximum values of $\sigma_{s,t+1}$, $\mu$ is the user-defined factor representing investor’s aversion to future risk. The smaller $\mu$ would lead risk constraints to be more sensitive to the fluctuation of trading performance in which a more strict risk requirement is assigned to trading agents in downtrend markets. The linear transformation derivation of $\sigma_{s}$ is described in the Appendix. Yet there may be no optimal solution to satisfy the strict risk constraint in terms of the current market situation. Thus, the risk constraint will be iteratively relaxed by a certain step size until the risk controller collects feasible solutions or meeting stop criteria.

\section{Experiments}

To carefully examine the performance of the proposed RiPO framework, the stock datasets with different market styles are selected to evaluate the RiPO and compared methods. There are three concerned questions in Section \ref{sec:perform}.

\subsection{Experimental Settings}

\textbf{Datasets}: To evaluate the performance of methods in the real financial market, the daily OLHCV data of constitute stocks of S\&P500 index in the U.S. market is collected from \textit{\href{https://finance.yahoo.com}{Yahoo Finance}}. The top 10 stocks are selected to construct a portfolio in terms of the company capital of stocks. The top 10 stocks, accounting for over 26\% of the S\&P500’s market capital, reflect U.S. market trends and provide high liquidity, satisfying the turnover rate assumption. Considering that most of the portfolio optimization methods may fail at the changes of market styles due to the data distribution shifts,  all compared methods will be tested on two market style settings. As defined in Table \ref{tab:1}, the three subsets of the MS-1 represent an uptrend financial market, aiming to compare the exploration ability of methods to search profitable strategies. On the other hand, the training set of the MS-2 depicts a relatively stationary market, but the validation data and test data are in the highly volatile and downtrend market due to the COVID-19 pandemic, which evaluates the risk management ability of methods when meeting unexpected crises.

\begin{table}
\centering
\caption{Description of the Dataset}
\scalebox{0.95}{
\begin{tabular}{llll}
    \toprule
    Dataset & Training & Validation & Test  \\
    \midrule
    MS-1 & @2015-@2017 & @2018  & @2019 \\
    MS-2 & @2015-@2019  & @2020  & @2021-10/31/22 \\
    \bottomrule
    \multicolumn{4}{c}{@: from $01/01$ to $12/31$ this year}\\
    \hline
    \end{tabular}%
    }
\label{tab:1}%
\end{table}

\noindent
\textbf{Comparative Methods}: In terms of investment principles like follow-the-winner, pattern matching, and DL/RL, seven methods are selected to compare with the proposed framework in this paper. They are Constant Rebalanced Portfolio (CRP, Neutral) \cite{cover1991universal}, Exponential Gradient (EG, Follow-the-Winner) \cite{helmbold1998line}, Online Moving Average Reversion (OLMAR, Follow-the-Loser) \cite{li26line}, Passive Aggressive Mean Reversion (PAMR, Follow-the-Loser) \cite{li2012pamr}, Correlation-driven Nonparametric Learning Strategy (CORN, Pattern Matching) \cite{li2011corn}, Ensemble of Identical Independent Evaluators (EIIE, DL/RL) \cite{jiang2017deep}, portfolio policy network (PPN, DL/RL) \cite{zhang2020cost}, Relation-Aware Transformer (RAT, DL/RL) \cite{xu2021relation}, and original Twin Delayed DDPG (TD3, DL/RL) \cite{fujimoto2018addressing}. 

\noindent
\textbf{Metrics}: To evaluate the returns and risks of compared methods, there are three common metrics in both academia and industry applied to measure the performance of trading strategies:

\begin{enumerate}[label=\arabic*]
    \item Annual Return: $\text{AR}=((1+\text{TR})^{\frac{252}{T}}-1)$, where $\text{TR}$ is the returns over the trading period and $T$ is the number of trading days.
    \item Maximum Drawdown: $\text{MDD}= \mathop{\max}\limits_{t_1 <t_2} \frac{C_{t_1}-C_{t_2}}{C_{t_1}}$, where $C_{t_1}$ and $C_{t_2}$ are the portfolio value at time $t_1,t_2$.
    \item Sharpe Ratio: $\text{SR}=\frac{\text{AR}-r_f}{\sigma_v}$, where $\sigma_v$ is the strategy volatility in terms of daily returns, and $r_f$ is the risk-free rate. Particularly, the $\text{SR}$ is assigned to 0 when the annual return is negative.
\end{enumerate}

\noindent
\textbf{Implementation Details}: TD3 is one of the popular RL algorithms and is selected to train RL-based trading agents in the RiPO framework. The implementation of the TD3 algorithm are referred by \cite{raffin2021stable} while the other baseline algorithms use default settings according to their papers. The detailed RiPO settings of experiments are given in the Appendix. Besides, for the fair comparison and avoiding future data leakage, the validation set are used to fine-tune the hyper-parameters while the test set is applied to compare the performance of methods. Furthermore, all experiments are repeatedly run for 10 times to avoid the randomness and bias of experiments caused by seeds. The average value and standard deviation of all methods are compared. Additionally, the Wilcoxon rank-sum test \cite{derrac2011practical} is used to compare the statistical significance of the proposed framework against the compared approaches with a significance level at 0.05.

\subsection{Performance Comparison and Analysis}\label{sec:perform}

\begin{figure}
    \centering
    \includegraphics[width=8cm]{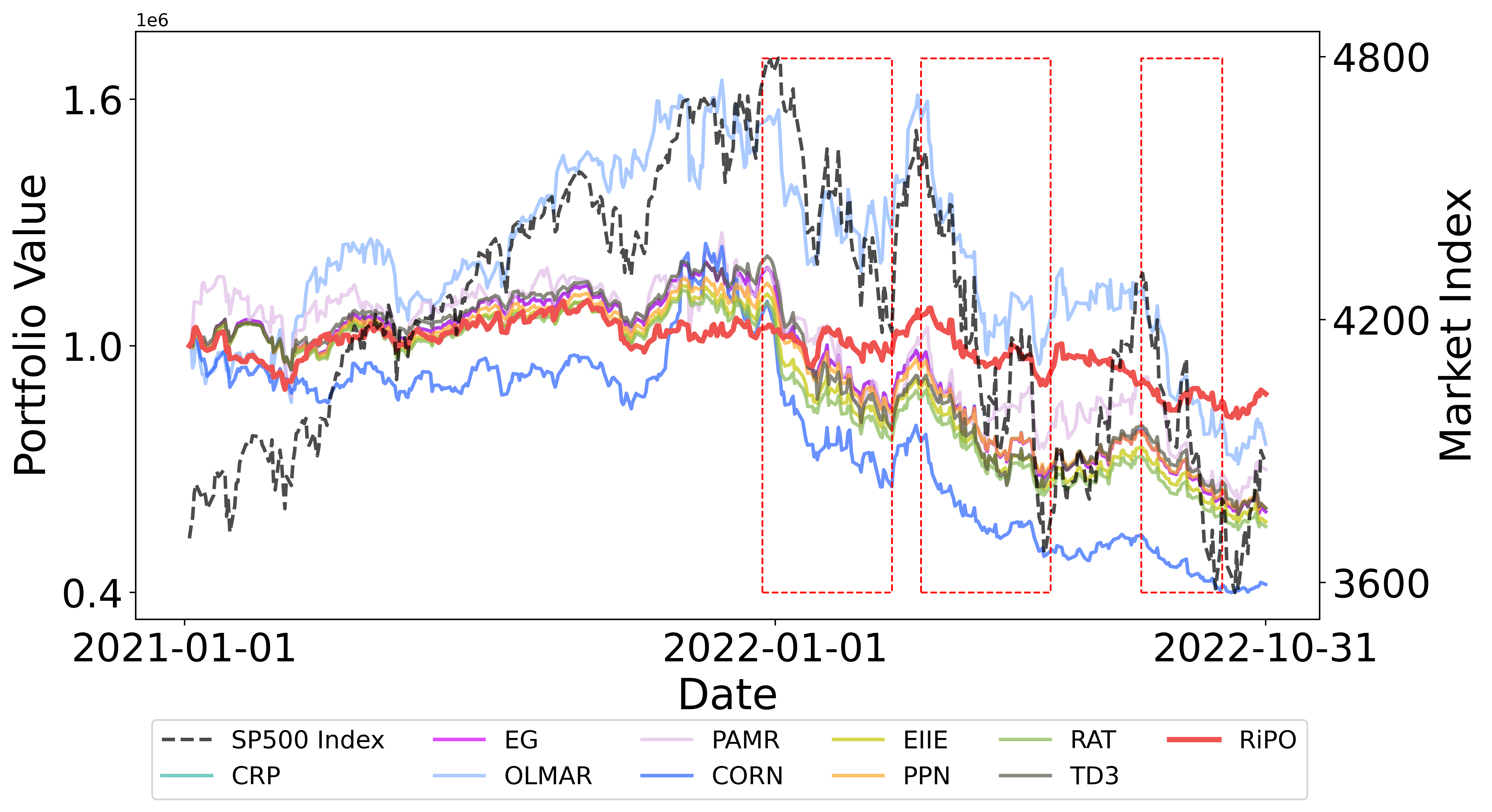}
    \caption{The Portfolio Value Comparison in the MS-2 Dataset.}% The results of ERM are from~\cite{iwasawa2021T3A}.}
    \label{fig:capital}
\end{figure}

\begin{figure}
    \centering
    \includegraphics[width=8cm]{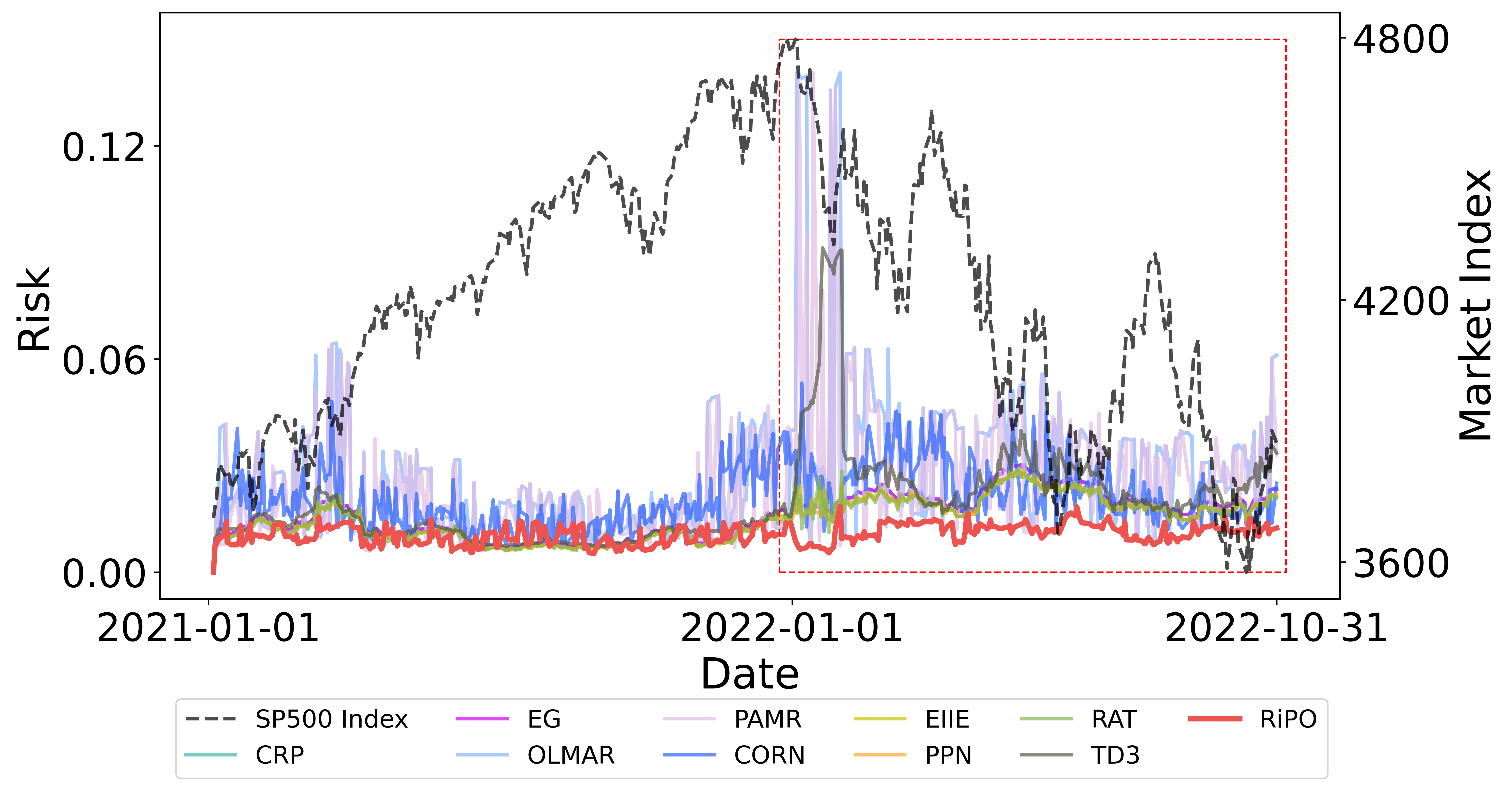}
    \caption{The Short-term Risk Comparison in the MS-2 Dataset.}% The results of ERM are from~\cite{iwasawa2021T3A}.}
    \label{fig:risk}
\end{figure}

\begin{table*}[ht]
  \centering
  \caption{Performance Comparison in Two Market Styles}
    \begin{tabular}{cccc|ccc}
    \hline
    % \multirow{2}{*}{\textbf{Metrics}} 
    & \multicolumn{3}{c|}{\textbf{MS-1 (Uptrend Market)}}  &\multicolumn{3}{c}{\textbf{MS-2 (Downtrend Market)}} \\
    \textbf{Method} & \textbf{AR$(\%)\uparrow$}     & \textbf{MDD$(\%)\downarrow$}     & \textbf{SR$\uparrow$}     & \textbf{AR$(\%)\uparrow$}     & \textbf{MDD$(\%)\downarrow$}     & \textbf{SR$\uparrow$}  \\
    \hline

    CRP   & 12.08 ($\pm0.0032$) & 17.11 ($\pm0.0010$) & 0.56 ($\pm0.0170$) & -24.57 ($\pm0.0008$) & 50.89 ($\pm0.0007$) & 0 \\
    EG    & 11.75 ($\pm0.0027$) & 17.10 ($\pm0.0010$) & 0.55 ($\pm0.0147$) & -24.63 ($\pm0.0019$) & 51.00 ($\pm0.0014$) & 0 \\
    OLMAR & -16.24 ($\pm0.0030$) & 55.08 ($\pm0.0018$) & 0 & -13.88 ($\pm0.0055$) & 56.74 ($\pm0.0035$) & 0 \\
    PAMR  & -30.86 ($\pm0.0053$) & 50.20 ($\pm0.0038$) & 0 & -17.79 ($\pm0.0043$) & 51.75 ($\pm0.0030$) & 0 \\
    CORN  & -9.93 ($\pm0.0058$) & 21.64 ($\pm0.0042$) & 0 & -37.83 ($\pm0.0024$) & 68.01 ($\pm0.0018$) & 0 \\
    EIIE  & 9.87 ($\pm0.0155$) & 15.86 ($\pm0.0024$) & 0.48 ($\pm0.0848$) & -26.33 ($\pm0.0081$) & 50.24 ($\pm0.0068$) & 0 \\
    PPN   & 12.48 ($\pm0.0188$) & \textbf{15.14 ($\pm0.0081$)} & 0.59 ($\pm0.1010$) & -23.98 ($\pm0.0038$) & 48.76 ($\pm0.0064$) & 0 \\
    RAT   & 12.26 ($\pm0.0315$) & 15.56 ($\pm0.0124$) & 0.60 ($\pm0.1651$) & -27.15 ($\pm0.0144$) & 51.41 ($\pm0.0192$) & 0 \\
    TD3   & 15.69 ($\pm0.1925$) & 27.65 ($\pm0.0902$) & 0.62 ($\pm0.3386$) & -24.26 ($\pm0.0660$) & 53.39 ($\pm0.0853$) & 0 \\
    \textbf{RiPO}  & \textbf{20.15 ($\pm0.1398$)} & 22.58 ($\pm0.0416$) & \textbf{0.72 ($\pm0.5138$)} & \textbf{-6.58 ($\pm0.0055$)} & \textbf{25.77 ($\pm0.0049$)} & 0 \\

    \hline
    \multicolumn{7}{l}{\small $\bullet$ Average value ($\pm$Standard deviation)} \\
    \end{tabular}%
  \label{tab:2}%
\end{table*}

\noindent
\textbf{Q1: How does the RiPO framework perform in terms of profitability and risk exposure in uptrend and downtrend markets?}

As shown in Table \ref{tab:2}, there are seven approaches achieving positive returns in the MS-1 dataset, among which the returns of the RiPO framework are at least $5\%$ higher than other methods while performing relatively low $\text{MDD}$ at around $22\%$. Although some baseline algorithms like PPN and RAT have lower $\text{MDD}$, their profitability is limited and some profitable strategies may be missed. The best $\text{SR}$ achieved at $0.72$ by the RiPO method demonstrates the capability to balance the profits and risk exposures of trading strategies. It reveals that the RiPO framework would dynamically relax the risk constraint within an acceptable range to pursue higher profits in uptrend markets. When testing in the MS-2 dataset, none of these baseline methods are profitable during the trading period. They may lose around $13\%$ to $37\%$ of the portfolio value for each year. Meanwhile, those methods have higher short-term risks in which investors may suffer a huge loss in a short period. Compared with baseline methods, the RiPO framework only loses $6.58\%$ for each year while the $\text{MDD}$ can be significantly reduced to $25.77\%$ against the $\text{MDD}$ of other methods that are over $48\%$. It demonstrates the outstanding ability of the RiPO to manage the risk in downtrend markets for avoiding great losses. Especially when encountering unexpected crises, the trading agents may not suggest reasonable trading signals where the risky investment should be strictly constrained. Besides, the significant test results $(better/equal/worse)$ are $4/5/0$ and $9/0/0$ in the MS-1 and MS-2, respectively. It is evident that the RiPO achieves remarkable performance in the MS-2 when compared to other baselines.

More importantly, the RiPO framework integrating risk controllers and TD3 as trading agents outperforms the original TD3 approach both in the MS-1 and MS-2 markets in terms of $\text{AR}$, $\text{MDD}$, and $\text{SR}$. This clearly demonstrates the effectiveness of the proposed risk controller to manage the potential investment risks due to the aggressive trading strategies generated by RL-based agents.

\noindent
\textbf{Q2: Can the RiPO framework effectively reduce downside risk?}

The downside risk is one of the most concerns for investors to manage portfolios. Fig. \ref{fig:capital} and Fig. \ref{fig:risk} show the portfolio value and short-term risk of compared methods in the test period. As highlighted in the red rectangular boxes, although the S\&P500 (black dash line) has a significant decline, the risky investments of the RiPO framework (red line) are strictly constrained to avoid huge losses while the portfolio value of other frameworks suffers a great loss. Besides, the short-term risks of the RiPO are carefully managed at a very low level than that of other methods during the whole downtrend period (see the red rectangular in Fig. \ref{fig:risk}), which further proves the outstanding capability of the RiPO to manage investment risks.

\noindent
\textbf{Q3: How do hyper-parameters that reflect the risk appetite of investor affect the RiPO framework?}

As described in the previous sections, there are three key hyper-parameters of the adaptive mechanisms DCM and ARS reflecting investors’ preferences to balance the strength of risk management and the exploration of trading strategies. As shown in Table \ref{tab:3}, the higher $m$ performs better $\text{AR}$ (from $-9.43\%$ to $-6.58\%$) and lower $\text{MDD}$ (from $29.47\%$ to $25.56\%$) by enhancing the impact of risk controller to the RL-based agents in downtrend markets. However, there may miss some profitable opportunities when $m$ is set to a high value (i.e., $m=1$). The $\text{AR}$ decreases to $-6.69\%$ as most risky investments are restricted. Similarly, the lower $v$ represents the higher risk aversion, which reduces the $\text{MDD}$ from $32.63\%$ to $25.77\%$ and avoids half of the loss when $v=0.005$ by tightening risk exposures. Moreover, using the appropriate scaling factor $\mu$ at $3$ in the MS-2 encourages the RiPO framework to balance the expected returns and short-term risks by dynamically adjusting the strength of risk constraints.

% \mlcomment{TBC}

\noindent
\textbf{Q4: How do the adaptive mechanisms DCM and ARS impact the RiPO framework?}

To enhance the flexibility of risk management in different financial markets, the DCM and ARS are introduced to adjust the impact of risk controllers and the strength of risk constraints. Table \ref{tab:4} shows that the ARS significantly reduces the losses from $23.32\%$ to $6.69\%$ when without using the DCM and from $18.76\%$ to $6.58\%$ when integrating the DCM. Meanwhile, the risk exposure decreases to half of the setting without the ARS. On the other hand, the RiPO framework involving the DCM can capture more potential profitable opportunities to avoid greater losses in which the short-term risks are managed at lower or similar levels by considering the tradeoff between long-term returns and short-term risks. The experimental results clearly reveal that the two adaptive mechanisms encourage the RiPO model to dynamically adjust risk constraints for both managing risk exposures and exploring profitable strategies.

\begin{table}
  \centering
  \caption{Key Hyper-parameter Analysis in the MS-2 Dataset}
    \begin{tabular}{cccc}
    \hline
     Parameter & Setting & \textbf{AR$(\%)\uparrow$}     & \textbf{MDD$(\%)\downarrow$}  \\
    \hline 

\multirow{5}{*}{$m$}     & 0     & -9.43 ($\pm0.0174$) & 29.47 ($\pm0.0270$) \\
                       & 0.2   & -8.51 ($\pm0.0148$) & 28.14 ($\pm0.0197$) \\
                       & 0.5   & -8.19 ($\pm0.0129$) & 28.07 ($\pm0.0138$) \\
                       & 0.8   & -6.58 ($\pm0.0055$) & 25.77 ($\pm0.0049$) \\
                       & 1.0     & -6.69 ($\pm0.0054$) & 25.56 ($\pm0.0081$) \\
                       \hline
\multirow{4}{*}{$v$}     & 0.005 & -6.58 ($\pm0.0055$) & 25.77 ($\pm0.0049$) \\
                       & 0.010  & -6.89 ($\pm0.0070$) & 26.02 ($\pm0.0111$) \\
                       & 0.100   & -7.11 ($\pm0.0160$) & 25.93 ($\pm0.0211$) \\
                       & 0.500   & -11.66 ($\pm0.0177$) & 32.63 ($\pm0.0278$) \\
                       \hline
\multirow{3}{*}{$\mu$} & 1     & -6.58 ($\pm0.0055$) & 25.77 ($\pm0.0049$) \\
                       & 2     & -7.35 ($\pm0.0051$) & 26.56 ($\pm0.0048$) \\
                       & 3     & -6.56 ($\pm0.0074$) & 25.77 ($\pm0.0125$) \\

    \hline
    \multicolumn{4}{l}{\small $\bullet$ Default settings: $m=0.8$, $v=0.005$, and $\mu=1$}\\
    \end{tabular}%
  \label{tab:3}%
\end{table}

\begin{table}
  \centering
  \caption{Ablation Studies on the Contribution of DCM and ARS Modules}
    \begin{tabular}{ccc}
    \hline
     Setting & \textbf{AR$(\%)\uparrow$}     & \textbf{MDD$(\%)\downarrow$}    \\
    \hline
        w/o ARS \& w/o DCM & -23.32 ($\pm0.0550$) & 48.85 ($\pm0.0727$) \\
        w/o ARS        & -18.76 ($\pm0.0387$) & 44.38 ($\pm0.0584$) \\
        w/o DCM        & -6.69 ($\pm0.0054$) & 25.56 ($\pm0.0081$) \\
        \textbf{RiPO}     & -6.58 ($\pm0.0055$) & 25.77 ($\pm0.0049$) \\
    \hline
    \end{tabular}%
  \label{tab:4}%
\end{table}

\section{Conclusion}

In the paper, a novel risk-manageable portfolio optimization framework named RiPO is proposed explicitly to manage the short-term risks while expecting the long-term profits in different market styles. With the cooperation of RL approaches and barrier function based risk controllers, the RiPO performs strong exploration ability to optimize trading strategies under acceptable risk constraints. Besides, two dynamic modules are given to construct a flexible risk controller for adapting financial markets and investors' risk appetite. The experimental results indicate that the RiPO can gain higher profits in uptrend markets and manage downside risks in downtrend markets. In the future, the flexibility of risk controller can be further enhanced for adapting to different financial markets and handling more real market constraints.

\bibliography{RiPO}

% #################################################
% Appendix
\clearpage

\appendix
\section{Technical Indicators}
Except for the open/high/low/close price and volume data, more advanced financial indicators are provided as market states in the RiPO framework to discover underlying trend patterns. All involved technical indicators are listed as below with the format \{Indicator name\}-\{Observation window size\}.

\noindent
AD: Chaikin A/D Line \\
ADOSC: Chaikin A/D Oscillator \\ 
ADX: Average Directional Movement Index \\
ADXR: Average Directional Movement Index Rating \\
APO: Absolute Price Oscillator  \\
AROON: Aroon \\
AROONOSC: Aroon Oscillator  \\
ATR-6/14: Average True Range-6/14 \\
BBANDS: Bollinger Bands   \\
BOP: Balance Of Power  \\
Change-open/high/low/close/volume \\
CCI-5/10/20/88: Commodity Channel Index-5/10/20/88 \\
CMO-open/close-14: Chande Momentum Oscillator-open/close-14  \\
DEMA-6/12/26: Double Exponential Moving Average-6/12/26 \\
DX: Directional Movement Index  \\
EMA-6/12/26: Exponential Moving Average-6/12/26  \\
KAMA: Kaufman Adaptive Moving Average  \\
MA-close-5: Moving Average-close-5 \\
MACD: Moving Average Convergence/Divergence \\
MEDPRICE: Median Price \\
MiNUSDI: Minus Directional Indicator \\
MiNUSDM: Minus Directional Movement  \\
MOM: Momentum  \\
NATR: Normalized Average True Range  \\
OBV: On Balance Volume  \\
PLUSDI: Plus Directional Indicator  \\
PLUSDM: Plus Directional Movement  \\
PPO: Percentage Price Oscillator \\
ROC-6/20: Rate of change-6/20   \\
ROCP-6/20: Rate of change Percentage-6/20 \\
ROC-volume-6/20: Rate of change-volume-6/20 \\
ROCP-volume-6/20: Rate of change Percentage-volume-6/20  \\
RSI: Relative Strength Index  \\
SAR: Parabolic SAR  \\
TEMA-6/12/26: Triple Exponential Moving Average-6/12/26 \\
TRANGE: True Range  \\
TYPPRICE: Typical Price \\
TSF: Time Series Forecast  \\
ULTOSC: Ultimate Oscillator \\
WILLR: Williams' \%R  \\

\section{Derivation of the Risk Controller}

The system dynamics and the risk controller are described in Section 4 of the paper, of which the barrier function-based risk constraint can be expressed as 

     \begin{equation}
        \begin{aligned}
        h\left(\sigma_{p, t+1}\right)-h\left(\sigma_{p, t}\right)+\alpha\left(h\left(\sigma_{p, t}\right)\right) \geq 0.
        \end{aligned}
    \end{equation}

To manage the investment risk within a given acceptable range, the portfolio risk $\sigma_{p, t}$ at time $t$ is constrained as the inequality:

     \begin{equation}
        \begin{aligned}
        0 \leq \sigma_{p, t} \leq \sigma_{s, t},
        \end{aligned}
    \end{equation}

\noindent where $\sigma_{s, t}$ is the upper boundary of the acceptable risk. Correspondingly, the barrier function $h$ can be defined as 

     \begin{equation}
        \begin{aligned}
        h\left(\sigma_{p, t}\right)=\sigma_{s, t}-\sigma_{p, t}.
        \end{aligned}
    \end{equation}

For the acceptable risk $\sigma_{p,t}\in[0,\sigma_{s,t}]$, there exists an acceptable set $C$ such that

     \begin{equation}
        \begin{aligned}
        C=\left\{\sigma_{p, t}: h\left(\sigma_{p, t}\right) \geq 0\right\}.
        \end{aligned}
    \end{equation}

Considering an extended class $\mathcal{K}_{\infty}$ function $\alpha\left(h\left(\sigma_{p, t}\right)\right)=\eta h\left(\sigma_{p, t}\right)$, $\eta \in[0,1]$, and the portfolio risk consists of the strategy risk $\sigma_{\alpha, t}$ and market risk $\sigma_{\beta, t}$, then the risk constraint can be further rewritten as

     \begin{equation}
        \begin{aligned}
        \sigma_{s, t+1}-\sigma_{\alpha, t+1}-\sigma_{\beta, t+1}+(\eta-1)\left(\sigma_{s, t}-\sigma_{\alpha, t}-\sigma_{\beta, t}\right) \geq 0.
        \end{aligned}
    \end{equation}

Furthermore, with the consideration of using an adaptive risk strategy in the RiPO framework, the expected short-term risk $\sigma_{p, t+1}$ at time $t+1$ can be managed within the acceptable range. If there exists an adjusted action $a_t^{Ctrl}$ from the risk controller, such that 

 \begin{equation}
    \begin{aligned}
    \sigma_{\alpha,t+1}&=\sqrt{a_t^T\ \Sigma_{k,t+1}a_t}\\
    &=\Vert \Sigma_{k,t+1}^{\frac{1}{2}} a_t\Vert_ 2 \\
    &= \Vert \Sigma_{k,t+1}^{\frac{1}{2}} a_t^{Ctrl}+\Sigma_{k,t+1}^{\frac{1}{2}}  a_t^{RL}\Vert_ 2 \\
    & \leq \sigma_{s, t+1}-\sigma_{\beta, t+1}+(\eta-1)\left(\sigma_{s, t}-\sigma_{\alpha, t}-\sigma_{\beta, t}\right) \\ 
    & \leq \sigma_{s, t+1}-\sigma_{\beta, t+1}+(\eta-1)(\sigma_{s, t}- \\ &\left\|\Sigma_{\mathrm{k}, \mathrm{t}}^{\frac{1}{2}} a_{t-1}\right\|_{2}-\sigma_{\beta, t}).
    \end{aligned}
\end{equation}

Let $A=\Sigma_{\mathrm{k}, t+1}^{\frac{1}{2}}$, $b=\Sigma_{\mathrm{k}, t+1}^{\frac{1}{2}} a_{t}^{R L}$, $c^{*}=\sigma_{s, t+1}-\sigma_{\beta, t+1}+(\eta-1)\left(\sigma_{s, t}-\left\|\Sigma_{\mathrm{k}, \mathrm{t}}^{\frac{1}{2}} a_{t-1}\right\|_{2}-\sigma_{\beta, t}\right)$, and $d=0$. Referred by the definition of second-order cone programming (SOCP) of the form $\|A x+b\|_{2} \leq d^{T} x+c^{*}$, the barrier function-based risk controller can be described as a SOCP problem with the linear objective function.

\section{Derivation of the Linear Transformation in Adaptive Risk Strategy}
The Adaptive Risk Strategy (ARS) module is introduced to adjust the upper boundary $\sigma_{s,t+1}$  of acceptable risk ranges. Too small $\sigma_{s,t+1}$ leads to the risk controller being unsolvable while too large $\sigma_{s,t+1}$ may not efficiently manage investment risks. Therefore, considering the current financial market states and the risk appetite of investors, there are three cases to adjust the acceptable risks $\sigma_{s,t+1}$. When the expected return $\bar{R}_{t+1}$ is lower than the user-defined lower boundary $(1-\mu)r_{f}$, the $\sigma_{s,t+1}$ will not decline and be fixed at $\sigma_{s, min}$. On the other hand, the $\sigma_{s,t+1}$ will be fixed at $\sigma_{s, max}$ for avoiding too large values when $\bar{R}_{t+1}$ exceeds the user-defined upper boundary $(1+\mu)r_{f}$. Lastly, when the $\bar{R}_{t+1}  \in\left[(1-\mu) r_{f},(1+\mu) r_{f}\right]$, a linear transformation is designed for the adjustment of $\sigma_{s,t+1}$. Define the linear transformation function $\sigma_{s,t+1} = M\bar{R}_{t+1} + b$, ($M, b \in\mathbf{R}$). There are two points $((1-\mu) r_{f}, \sigma_{s, min})$ and $((1+\mu) r_{f}, \sigma_{s, max})$ on the linear function. Then, the following two equations can be obtained.

    \begin{equation}
        \begin{aligned}
            \left\{\begin{array}{l}
            \sigma_{s, \min }=M(1-\mu) r_{f}+b \\
            \sigma_{s, \max }=M(1+\mu) r_{f}+b
            \end{array}\right.
        \end{aligned}
    \end{equation}
    
\noindent
where $\sigma_{s,min}$ and $\sigma_{s,max}$ are the minimum and maximum values of $\sigma_{s,t+1}$, $\mu$ is the user-defined factor representing investor’s aversion to future risk, and $r_{f}$ is the risk-free rate. After solving the above two equations, it has 
    \begin{equation}
        \begin{aligned}
            \left\{\begin{array}{l}
            M=\frac{\sigma_{s, \max }-\sigma_{s, \min }}{2 \mu r_{f}} \\
            b=\frac{(1+\mu) \sigma_{s, \min }-(1-\mu) \sigma_{s, \max }}{2 \mu}
            \end{array}\right.
        \end{aligned}
    \end{equation}

Lastly, when $\bar{R}_{t+1}  \in\left[(1-\mu) r_{f},(1+\mu) r_{f}\right]$, the linear transformation function can be described as below.
    \begin{equation}
        \begin{aligned}
            \sigma_{s}=\frac{\sigma_{s, \max }-\sigma_{s, \min }}{2 \mu r_{f}} \bar{R}_{t+1}+\frac{(1+\mu) \sigma_{s, \min }-(1-\mu) \sigma_{s, \max }}{2 \mu}.
        \end{aligned}
    \end{equation}  

\section{Experimental settings}

In the RiPO framework, the TD3 method is integrated into the RiPO for training the RL-based trading agents. More specifically, the policy network is a two-layer architecture with $400$ and $300$ hidden units, respectively. The learning rate is $10^{-5}$, and the memory size is $10^6$. The RL model is updated every $400$ steps with the $50$ batch size. The training episode of RL algorithms is $500$. Besides, the policy and target networks are updated every $2$ steps per training step, and the standard deviation of Gaussian noise added to the target policy is $0.2$. Additionally, the $\eta$ in an extended class $\mathcal{K}_{\infty}$ function $\alpha$ is set to $0.3$. To balance the returns and risks, the risk appetite of investors $v$ and the minimal impact of risk controllers $m$ in the dynamic contribution mechanism default to $v=0.5$ and $m=0$ in the MS-1 dataset, and $v=0.005$ and $m=0.8$ in the MS-2 dataset. In the adaptive risk strategy, the scaling factor $\mu$ is set to 2 in the MS-1 dataset and 1 in the MS-2 datasets. The minimum and maximum values of $\sigma_{s, t+1}$ are set to $\sigma_{s,min}=0.01$ and $\sigma_{s,max}=0.02$ in the MS-1 dataset, and $\sigma_{s,min}=0.01$ and $\sigma_{s,max}=0.015$ in the MS-2 dataset. Moreover, in terms of different market styles, the window sizes for observing recent strategy performance to estimate the expected returns are 3 days and 5 days for MS-1 and MS-2 datasets, respectively.

Furthermore, to simulate the trading in real markets, the risk-free rate is $1.6575\%$ referred by the U.S. 5-year bond yield. Initially, the capital is $1000000$, and the market risk is assumed to $\sigma_\beta=0.001$. Also, the transaction cost is set to $0.1\%$, while the slippage defaults to $0.1\%$. Moreover, the observation window size of covariance calculation is $21$ days (i.e., a month) based on the usual practice of investors. For the calculation of the yearly performance indicators, the number of trading days for a year defaults to $252$. All algorithms are implemented by Python on a desktop computer installed with the 2 NVIDIA RTX 3090 GPUs, and all experiments are run for 10 times to avoid the bias of results caused by random seeds.  

\section{Implementation Code of the RiPO Framework}

The implementation code of the RiPO framework and the S\&P500 market data are attached in the supplemental materials. Please contact the authors for the implementation.

% \ack We would like to thank the referees for their comments, which
% helped improve this paper considerably

\end{document}